Dejan Grba[1]
Digital Art Program, Interdisciplinary Graduate Center, University of the Arts in Belgrade
dejan.grba@gmail.com

**The Mechanical Turkness: Tactical Media Art and the Critique of Corporate AI**

**Abstract**

The extensive industrialization of artificial intelligence (AI) since the mid-2010s has increasingly motivated artists to address its economic and sociopolitical consequences. In this chapter, I discuss interrelated art practices that thematize creative agency, crowdsourced labor, and delegated artmaking to reveal the social rootage of AI technologies and underline the productive human roles in their development. I focus on works whose poetic features indicate broader issues of contemporary AI-influenced science, technology, economy, and society. By exploring the conceptual, methodological, and ethical aspects of their effectiveness in disrupting the political regime of corporate AI, I identify several problems that affect their tactical impact and outline potential avenues for tackling the challenges and advancing the field.

---

[1] Corresponding author

**Introduction**

Since its largely obscure beginnings in the 1970s, AI art has expanded, gained visibility, and attained cultural relevance since the second half of the 2010s (Audry, 2021). It has benefited from the accelerating affordance of multilayered subsymbolic machine learning (ML) architectures such as Deep Learning, AI's increasing sociopolitical impact, and the art market's integration with the crypto economy. Contemporary AI art includes diverse creative pathways informed by functions, applications, and consequences of modern ML systems (Cetinić and She, 2022; Grba, 2022a, 3–17).

The rapid industrialization of AI drives the massive concentration of wealth, power, and (often controversial) sociopolitical influence, which define a space for the critical discourse that traces and questions the constitution of social reality through the technical logic of AI. Artists play a significant part in that engagement. They deconstruct, reassess, reappropriate, and sometimes democratize AI technologies to uncover or underline AI development's epistemic and existential issues and question the economic and sociopolitical consequences of AI capitalism. Their production continues the flux of heterogeneous tactical media practices that have energized art and culture since the late 20th century with hybrid forms of academic or critical interventions into the neoliberal condition's technological, political, economic, and cultural layers. With the increasing accessibility of technologies that can be modified and repurposed by the actors who operate outside of the established hierarchies of power and knowledge, tactical media art has emerged as a response to a shift in postindustrial society towards the information economy in which efficiency, operationalism, and instrumental rationality become core values, and market transactions the predominant social good.

Although it often maps top-down power relations, tactical media art embodies a sense of bottom-up resistance in a manner and style associated with cultural dissent and opposition. It challenges the dominant semiotic regime through signs, messages, and narratives that foster critical thinking and offer new ways of seeing, understanding, and (in some scenarios) interacting with the systems of power. Tactical media works are not sweeping revolutionary events but micro-political acts of disruption, intervention, and education for which absolute victory or fundamental transformation is neither a desirable nor truly attainable objective. They usually generate fleeting, ephemeral, and pliable statements or initiatives that must be continually reconfigured in response to their changing targets. Their effects are often not immediate but cumulative and relational because they provide insights, knowledge, and tools that may become transformative in the audience's hands (Raley, 2009).

In different ways, tactical media art contends with business or state strategies based on quantization, statistical reductionism, datafication, behavioral prediction, and the manipulation of decision-making (Grba, 2020, 71–73). AI technologies increase these strategies in scope, extent, and sophistication, often amplifying their undesirable consequences. Hence, a notable portion of contemporary tactical art's topics involves biases, prejudices, inequities, and political agendas in the AI architectures and relates to the AI industry's exploitative practices. Artists critique the aggregation of transnational labor for training and testing ML models, the systematic concealment of the extent, value, and implications of human creative endeavor in AI development, and the immobilization of the AI workforce through manipulative interfaces, complacency, lack of protection, and precarity. Their expressive methodologies do not always follow overtly activist agendas but offer subtle, sometimes covert or intentionally ambivalent, critical narratives. They occasionally combine provocation with humor by recontextualizing AI systems' objectives or operative modes for ironic revelatory effects. Successful works usually let the audience actively identify corporate AI's interests, animosities, and injustices. They emphasize presence, engagement, and response but sometimes constrain them to raise the audience's awareness of limited actionability or immediate political impact. The tactical AI art's critical value is proportional to the artists' understanding of the political, economic, and moral nuances dispersed across the abstract, technically convoluted, and functionally opaque realities of AI technologies and their applications. This makes tactical AI art conducive to understanding how AI simultaneously reflects and influences sociopolitical relations, economies, and worldviews.

The existing literature pertinent to tactical AI art includes Marcus and Davis' *Rebooting AI* (2019), Mitchell's *Artificial Intelligence* (2019), and Larson's *The Myth of Artificial Intelligence* (2021), which pinpoint conceptual and technological issues of AI research and contradictions of AI implementation. Pasquinelli's *How a Machine Learns and Fails* (2019) and Kearns and Roth's *The Ethical Algorithm* (2019) address the ethical and sociopolitical consequences of AI's technical imperfections and biases. Żylińska's *AI Art* (2020) critiques AI's impact on art and culture. Crawford's *Atlas of AI* (2021) maps the exploitative layers of AI capitalism hidden behind marketing, media hype, and cultural commodification. In *Tactical Entanglements* (2021), Zeilinger provides a theoretical analysis of selected AI artworks' tactical values, issues, and potentials. In *Deep Else* and *Lures of Engagement* (both 2022) and *Faux Semblants* and *Renegade X* (both 2023), I critique the creative, expressive, and ethical features of AI art.

My discussion in this chapter centers around art practices that thematize creative agency, crowdsourced labor, and delegated artmaking to reveal the social rootage of AI technologies, underline productive human roles behind the performative power of applied AI, and expose conflicted social relations in the corporate AI sector. I focus on works whose poetics indicate broader issues of contemporary AI-influenced science, economy, and society. Lined out in three subject areas—The Elusive Artist, Digital Labor Transparency, and Outsourcing Artistic Work—these exemplars are interlinked across their topical domains and art-historical contexts.[2] I examine the conceptual, methodological, expressive, and ethical aspects of these works and assess their effectiveness in impacting AI's phenomenological, epistemic, and political realities. This allows me to identify several problems that affect their tactical cogency and to outline some prospects for addressing the challenges and advancing the field.

## Human Shades of AI

The increased enterprise and state application of AI has a significant influence on culture, economy, and politics. Many problematic aspects of that influence are seductive and hard to assess because of strong anthropomorphic[3] concepts in the foundations of AI science, AI's convoluted technical logic, and the IT industry's systemic illusionism of autonomous AI (Larson, 2021; Natale, 2021). Their impacts on artists range from provoking critical scrutiny, through inciting confusion, to inducing complacency.

---

[2] All discussed works are well documented and included in the References, so I compacted their descriptions to the topically most pertinent aspects. Details of additional listed exemplars can be found online by querying artists' names and work titles.

[3] Anthropomorphism is an innate psychological tendency to assign human cognitive traits, emotions, intentions, or behavioral features to non-human entities or phenomena (Hutson 2012).

*The Elusive Artist*

The specter of misidentified creative autonomy has been haunting computational art since its outset because the pervasiveness and efficacy of computing technologies can easily trick artists into conflating (cumulative) human creativity with its highly formalized technical emulations. Pioneering AI artist Harold Cohen had an ambiguous relationship with the machinic creative agency in his lifelong project called *AARON* (1971–2016)—a robotic system tasked to "draw and paint autonomously" and "embody creative behavior and the conjuring of meaning" (Cohen, 1995). Cohen followed computationalist doctrines and iteratively programmed the robot to generate images through the interaction of symbolically encoded cognitive primitives and render them on paper or canvas via different hardware interfaces (Taylor, 2014, 126–134). Whether sincerely or for promotional purposes, Cohen flirted with mystifying rhetoric about the robot's "surprises" and "creative serendipity" (Garcia, 2016), which converged with his pioneering role into a strong tributary to the legacy of bio-detached anthropomorphism in computational art.

The emotional charge of contemporary AI artists' claims that "there is something deeply thrilling about observing a machine learn, starting from scratch and iteratively discovering something about its world" (Audry, 2021, 85) indicates a strange fascination with complex statistical computation situated within strictly defined expressive spaces and signals an inclination to elevate constrained modes of functional autonomy into meaningful cognitive processes. Rather than fundamentally approaching their AI applications as tools, artists frequently represent them as "autonomous creators", "creative collaborators", "partners", or "companions" (Audry, 2021, 27–28, 241–243).[4] Despite repeated criticism and debunking (Browne and Swift, 2019; Browne, 2022), this uncritical rhetoric has steadily gained momentum among artists and has been assimilated into art scholarship.[5]

Some artists exploit anthropomorphic impulses through saccharine repetitions of Cohen's emphasis on robotic creative agency and misleading claims that they "teach their robots how to draw or paint" instead of using dryer but more accurate language about constructing and programming robotic systems to produce drawings and paintings. Examples include Pindar Van Arman's *Painting Robots* (since 2006) (Van Arman, 2016), Shantell Martin and Sarah Schwettmann's *Mind the Machine* (2017) (Schwettmann, 2017), and Joanne Hastie's *Abstractions (Tech Art Paintings)* (since 2017) (Hastie, 2021). Driven by weekend artists' enthusiasm, these works symbiose the happy-go-lucky joy in technocentric creativity with wafer-thin conceptualization and dilettante negligence toward the evolution of visual and media arts since the late 19th century.[6] They collapse the meaning of "art" into banal algorithmically rendered visualizations, self-confidently presented[7] in a series of "progressively improved" ML systems.

---

[4] Mainstream AI-derivative art is replete with anthropomorphizations of ML systems. For instance, Holly Herndon calls her vocal deepfake generator Holly+ her "digital twin" and pronounces it as "she/her" (Herndon 2022). In a statement about her and Jlin's work *Godmother* (2018), Herndon calls the software they used the "nascent machine intelligence" that "listened to her godmother [Jlin]" and reinterpreted her (Jlin's) art in the voice of "her mother [Herndon]" (Audry 2021, 213).

[5] See, for instance, the language used to announce the thematic scope and interest areas of the 2024 ISEA symposium (ISEA 2023).

[6] See, for example, Hopkins (2000) and Arnason and Mansfield (2012).

[7] For instance, art critic Jerry Saltz disparaged Van Arman's and several other AI artists' works for aesthetic dilettantism, conceptual shallowness, and ignorance of art historical context in the AI Art Reviews episode on Vice News. However, following the wisdom that every publicity is beneficial, Van Arman opens the press section on his

But *Ai-Da Robot Artist* (2019), developed by gallerist Aidan Meller and a Cornish robotics company Engineered Arts (Meller et al., 2019), integrates the anthropomorphic AI kitsch perhaps most blatantly in its design, visual output, and press coverage.

Attractive to the audience and the media, simplistic or superficial takes on machinic creativity, authorship, and originality can be found in all areas of AI art. For instance, some performance artists who enjoy the support of big tech companies aid their sponsors' public relations campaigns by representing AI as a pantheon of powerful but friendly anthropomorphic deities. They imply notions of machinic creative spontaneity in human-robot interactions that exploit the evolved human capacity for, and bias toward, detecting agency in midsized objects moving at medium speeds. Their works promote a robotically enhanced consumerist lifestyle or muse about existentially intense but politically or ethically vague notions of human-AI symbiosis. They are also sleekly sanitized and anesthetized mutations of earlier avantgarde practices in media art, experimental music, theater, and performance. Examples include Huang Yi's choreography *HUANG YI & KUKA* (since 2015) featuring Yi, two more dancers, and a preprogrammed KUKA industrial robot in a series of vignettes that beautify "the sorrow and sadness of Huang's childhood; an expression of loneliness, self-doubt, self-realization, and self-comfort" (Yi, 2021); Sougwen Chung's *Drawing Operations* (since 2015)—a series of performances augmented by video projections and soundtracks, in which Chung interacts with several versions of robots named Drawing Operations Units (custom-designed in collaboration with Bell Labs) to produce drawings that thematize mimicry, memory, and future speculations (Chung, 2020); Nigel John Stanford's tour de force music video for the title track on his album *Automatica: Robots vs. Music* (2017) in which he and several preprogrammed KUKA robots play and ultimately destroy various instruments (Stanford, 2017); and visceral homo-robotic encounters such as Marco Donnarumma and Margherita Pevere's *Eingeweide* (2018)—a "ritual of coalescence" in which Pevere and Donnarumma, wearing robotic prostheses controlled by real-time biomimetic algorithms, violently perform on stage adorned with out-of-body organs that pulsate and leak traces of microbial cultures, relics from computer server farms, and animal remains in an immersive soundscape of the artists' muscular activity modified and amplified by another set of algorithms (Donnarumma, 2023).[8]

The audience's romantically skewed perception of, and virtue signaling about, AI art reinforces the anthropomorphization of AI in popular discourse. Georgina Ruff (2022) exemplified this trend by dissecting the disparities between the "emphatic" verbal exchanges about Sun Yuan and Peng Yu's robot installation *Can't Help Myself* (2016), its actual production background, and exhibition history. She identified and analyzed heavy anthropomorphization in both the exhibition curator's comments and in social media narratives communally constructed, circulated, and reverberated by the audience who largely never experienced the work in person. Such public feedback in turn encourages artists to "relegate" their creative agency to ML architectures. Recurring with each increase in AI technologies' precision or scope, the latest round of this hype cycle is the proliferation of prompt-generated images, videos, and animations with large Text-to-Image diffusion models, such as DALL-E 2, MidJourney, Stable Diffusion, Disco Diffusion, and Pytti (McCormack et al., 2023).

---

website by quoting the first part of Saltz's comment on his work: "It doesn't look like a computer made it…" (Van Arman 2023). The complete comment was: "This is the first image [in a sequence presented by the Vise team] I've seen that does not look like a computer made it. Robot really likes post-impressionism; that's a good taste. Could I mistake it for a real person's hand? Sure. Certainly, it does not make it better." (Kumar 2018).

[8] See the critique of these works in Grba (2022a, 5).

However, anthropomorphization also allows artists to employ critical thinking and contextual awareness to address the questionable intersections of AI and creativity. For instance, in Adam Basanta's installation *All We'd Ever Need Is One Another* (2018), custom software randomizes the settings of two mutually facing flatbed scanners so that in every scanning cycle, each captures a slightly altered mix of the facing scanner's light and its unfocused scanning light reflected off the facing scanner's glass plate. The perceptual hashing algorithms use parameters such as aspect ratio, composition, shape, and color distribution to compare each scan to a database of images and image metadata scraped from freely accessible online artwork repositories. When the comparison value between the scan and the most similar database image exceeds 83 percent, the software declares a "match", selects the scan for printing, and labels it according to the database image metadata (Basanta, 2018). After it printed one of the scans and labeled it *85.81%_match: Amel Chamandy 'Your World Without Paper', 2009*, Canadian artist Amel Chamandy initiated a legal action about the intellectual property rights against Basanta because of the reference to her photograph even though *85.81%_match…* is not for sale and Basanta does not use it for direct commercial gains by any other means. In *All We'd Ever Need…*, he legitimately and consistently applied the functional logic of ML to leverage the open-endedness of creative work and disturb the entrenched notions of agency, authorship, originality, and intellectual property. *85.81%_match…* provides a cue for argumentation intricacies in Chamandy's lawsuit to reveal the cognitive and ethical issues of our tendency to crystalize creativity with commercial rights (Zeilinger, 2021, 94–108).

Basanta and other notable artists criticize AI as a sociopolitical complex and as a technology. They expose the misalignment between algorithmic decision-making processes and their representative anthropomorphic metaphors and question the responsibility of using such metaphors in the AI's underlying epistemologies (Curry, 2023, 181). Recognizing the variable abstractability of technologically entangled authorship, they show that crucial aesthetic factors such as decision-making, assessment, and selection are human-driven and socially embedded regardless of the complexity or counterintuitiveness of our expressive tools. By emphasizing that artworks emerge and live in ceaseless interrelations with other creative acts and agencies (Zeilinger, 2021, 104), they reaffirm the notion of art as a dynamic, evolving, biologically and sociopolitically contextualized faculty that needs continuous cultivation. Although these properties make art hard to objectify, commercialize, and own, it remains instrumentalizable in myriad ways.

*Digital Labor Transparency*

The corporate AI sector integrates AI industry giants, such as Google, Amazon, Microsoft, IBM, Apple, and Meta in the USA and Baidu, Alibaba, Tencent, and Huawei in China, with numerous smaller enterprises, startups, affiliated academic programs, and independent researchers. Its business politics unfolds in defining and controlling research, development, marketing, and application of AI technologies. Both AI-producing and AI-powered companies deploy cybernetic frameworks to "optimize" labor, maximize profit, and reinforce their power structures. Continuing the capitalist paradigm of labor transparency, the AI industry simultaneously strives to obscure the economic interests behind its commodities and render its workforce invisible in professional and public discourse. One of the main strategies to achieve this is by concealing or diminishing the extent, value, and implications of human endeavor in AI development and foregrounding the notions of autonomous AI.

However, like Wolfgang von Kempelen's fake chess-playing automaton The Automaton Chess Player (also called The Mechanical Turk or The Turk),[9] enterprise AI applications are simulacra of machinic autonomy substantiated by a heterogenous corpus of computer scientists, software engineers, online workers, and everyday users. Online microlabor crowdsourcing platforms, such as Amazon's Mechanical Turk (MTurk),[10] are arguably the most notorious instances of corporate AI's "human-in-the-loop" complex. They have been extensively used as largely unregulated instruments for amassing transnational echelons of workers to perform repetitive tasks essential for building, training, and testing ML architectures. While online microlabor provides workers some new opportunities to renegotiate power relations, it was initiated by, and for the convenience of, the techno-capitalist elite to push labor extraction, division, alienation, and precarity to new extremes, thus entrenching or even exacerbating the established power differentials (Golumbia, 2015; Hill, 2016). The AI community adopted Amazon's cynical nomenclature to euphemistically call the online-aggregated human labor "pseudo-AI" or "Artificial Artificial Intelligence" (AAI).

Artists address this foundational "artificiality" by revealing corporate AI's sociopolitical conflicts and exposing the productive human roles in AI technologies. As a prescient critique of the 1990s hype about the "autonomy" of interactive digital multimedia, Mona Hatoum's three-day performance *Pull* (1995) provides a contextual backdrop for the "human in the loop" complex. *Pull*'s visitors were invited to pull the braid of human hair hanging in a wall niche and witness in real-time Hatoum's facial and vocal reactions on a video monitor directly above it. Fascinated by the speed and variability of mediated interaction, visitors largely perceived the work as an advanced digital installation, whereas it was based on a von Kempelen-style deception. The braid was not part of a computer interface but the natural hair of Hatoum, who had laid hidden in a chamber behind the wall and faced the live video camera/microphone setup (Artbasher, 2007). Matt Richardson's *Descriptive Camera* (2012) belongs to a related body of works that address the sociotechnical blindness of the AI economy. *Descriptive Camera* has a conventional lens and sensor but no display or memory; instead, it sends the captured image to a MTurk worker (MTurker) tasked to write down and upload its brief description back to the device which prints it out (Richardson, 2012). This "picture development" process offers a revelatory counterintuitive glimpse into the large volume of data classification and annotation tasks in building ML training datasets, which the AI industry outsources to underpaid workers worldwide.

Of course, the corporate AI's sociotechnical blindness does not manifest as a lack of perceptiveness for the centrality and values of human work but as an active evasion of acknowledging them. RyBN and Marie Lechner's project *Human Computers* (2016–2019) targets this evasiveness by documenting the historical use of human beings in large computational architectures (RyBN, 2021). It features an online media archaeology of human labor in computation since the 18th century, from von Kempelen's Mechanical Turk to recent enterprises such as SETI@home, MTurk, ReCPATCHA, Scale, etc. In 2018, this project included an online game *AAI Chess*, which offered three all-human playing modes: human vs. human, human vs. MTurker, and MTurker vs. MTurker. Two years later, Jeff Thompson used the same principle in his *Human Computers* (2020). It tasked visitors to manually resolve a digital image file (a Google StreetView screenshot of the gallery) from its binary form into a grid of pixels. With 67 calculations per pixel, the complete human-powered image assembly takes approximately eight hours (Thompson, 2020). Here, the visitors' enactment of automated computation illustrates how a combination of complexity and speed in pervasive technologies makes them efficient but hard to understand and manage.

---

[9] Wolfgang von Kempelen constructed the Turk in 1770. For 84 years, it was exhibited throughout Europe and the Americas as a fully automatic chess-playing machine, attracting many spectators and notable challengers, such as Napoleon Bonaparte, Charles Babbage, Benjamin Franklin, and Edgar Allan Poe (Carroll 1975). The Turk was designed as a torso of an Ottoman sultan fronting the chessboard mounted on top of a cabinet with an elaborate internal mechanism. However, the mechanism's role was not to compute the moves by analyzing chess table positions but to conceal the expert human chess player and enable him to move the sultan's figures on board (Jay 2001).

[10] Besides MTurk, online microlabor platforms include Fiverr, Microworkers, Clickworker, Upwork, TaskRabbit, WorkMarket, Catalant Technologies, Inc., Toloka, and others.

In *Crowd-Sourced Intelligence Agency (CSIA)* (since 2015), Derek Curry and Jennifer Gradecki address the related issues of accuracy and accountability in data processing. *CSIA* offers a vivid educational journey through problems, assumptions, and oversights inherent in ML-powered dataveillance. It centers around an online app that partially replicates an OSINT system and allows the visitors to assume the role of data security analysts by monitoring and evaluating their friends' Twitter messages or by using an automated Bayesian classifier to test the "delicacy" of their own messages before posting (Gradecki and Curry, 2017). This relational architecture fosters active learning by combining a library of resources about the intelligence sector's analytic and decision-making practices with playfully transgressive "message policing" that allows a direct experience of how the inaccurate judgment of decontextualized metadata affects ML systems used by the surveillance agencies.

Aggregated human input is crucial for the initial configuration of ML models but also for their refinement with real-world feedback data, in which content recommendation algorithms play an interesting part. Tuned in most cases for monetary or political gain, they retain the online platform users' attention and influence their actions by predicting the content a user will favor based on their previous interactions. Besides being manipulative by design, predictive algorithms remain notoriously mechanistic and solipsistic because they are tasked to shape complex human behavior based on relatively simplistic and decontextualized online data. Lacking commonsense reasoning capabilities, they also tend to be abusive, so if a user selects biased or disinformational content, the following recommendations may reinforce the slanted or false narrative. Tomo Kihara's online work *TheirTube* (2020) warns about the implications of such flaws. It is a filter-bubble simulator that generates YouTube landing pages with recommended videos for six fictional profiles (Liberal, Conservative, Conspiracist, Climate Denier, Fruitarian, and Prepper) based on a simulation of their previous YouTube viewing data (Kihara, 2020).

With *Randomized Living* (2015–2017), Max Hawkins undertook a riskier and perhaps more responsible interrogation of the applied AI's consequences. In this two-year experiment, Hawkins lived according to the dictate of recommendation algorithms. He designed several apps that continuously harvested his online data to suggest a city where he should reside, places to go there, people to meet, and things to do. Hawkins followed these suggestions for about a month in each recommended city (Hawkins, 2021). *Randomized Living* is a striking exemplar of cybernetic existentialism—the art of conceiving a responsive and evolving cybernetic system to express profound existential concerns. It warns about our tendency to constrain behavior and cognition to fit certain protocols, such as labor regimes, and our susceptibility to perceptual and expressive conditioning in interactions with computationally mediated reality (Pasquinelli, 2019, 17).

These works reveal the technological reductivism that models social relations by mutually reinforcing network effects and the online users' opportunism or complacency. Beyond corporate AI, technological reductivism defines the pathological business logic of many IT providers, which dehumanizes users and turns them into slavish data-generating commodities by inciting addiction to "free" online services whose recommendation algorithms tend to favor negatively biased, politically derisive, and socially toxic views. Although such deviations get somewhat compensated over time by the emergent flexibility of human intelligence that drives cultural maturation, they are nonetheless socially damaging because of the asymmetry in production means and political power between corporate AI institutions and individuals.

*Outsourcing Artistic Work*

Artmaking is always contextualized by the sociocultural environment and dependent on the available knowledge, material, and organizational resources, so delegating some of its components is inevitable. The outsourcing of artistic labor has a long history reaching back to antiquity. In Western culture, it is emblematized by Renaissance artists' studios that employed technicians, apprentices, and disciples to produce commissioned works with various degrees of the master artist's involvement (Wallace, 2014). Traditionally, outsourced creative labor remained concealed in the finished artwork but avantgarde modernist artists rendered it explicit by developing techniques such as collage, assemblage, and bricolage to experiment with the nature of creativity, question the cultural notions of art, and challenge the parameters of aesthetic judgment. Pablo Picasso, Georges Braque, Hannah Höch, Kurt Schwitters, and others legitimized creative outsourcing in their practices, and Marcel Duchamp radicalized it with readymades, such as the venerable *Fountain* (1917) (Molderings, 2010). Following the proliferation of artistic outsourcing strategies throughout the 20th century, the Internet facilitated their further expansion by coupling accessibility with programmability.

For instance, Perry Bard brilliantly engaged the relational flexibility of human visual interpretation in her online project *Man with a Movie Camera: The Global Remake* (2007–2014). It invited visitors to select any shot from Dziga Vertov's seminal film *Man with a Movie Camera* (1929) and upload their video interpretations (Bard, 2014). Bard's server-side software collected participants' shots and made a two-channel synchronized composition with Vertov's original playing on the left and its continuously reassembled remake on the right.[11] By leveraging the breadth of human perceptive cognition, this relatively simple technical setup engrosses both uploaders and viewers in an intriguing experience of surprise motivated by curiosity rather than monetary profit: in this non-commercial project, Bard used her webpage and software to transparently define, solicit, and process the visitors' voluntary participation based on a common-exploratory goal.

Since the launch of online microlabor platforms in 2005, artists have used them in diverse creative scenarios. For example, Clement Valla ran several generative experiments through MTurk. In *Sol LeWitt + Mechanical Turk* (2009), Valla's software recreated Sol LeWitt's algorithmic drawings, posted their instructions for MTurkers to execute online (5 US cents per drawing), and assembled their interpretations into a grid. In *A Sequence of Lines Traced by Five Hundred Individuals* and *A Sequence of Circles Traced by Five Hundred Individuals* (both 2011), Valla utilized the entropic effects of iterative tracing: both works were realized as online drawing tools that let users trace a line (*A Sequence of Lines…*) and a circle (A Sequence of Circles…) and chronologically recorded the on-screen tracing process. The first MTurker was shown a straight line/circle to trace but each successive one only saw the previously produced trace so the copying imperfections accumulated. MTurkers were paid 2 US cents per trace. In *Seed Drawing* (2011), the artist added evolutionary effects to the process of iterative copying: over a 3-month period, thousands of MTurkers were tasked to copy simple line drawings following the same rules as in the previous two works and their output was aggregated into a large-scale structure of organic patterns (Valla, 2023).[12]

---

[11] Multiple interpretations of the original shot were randomly selected for the replay.

Artists have also experimented with tactical arrangements of delegated creativity on microlabor platforms to highlight and question their socioeconomic issues (Bishop, 2012; Torok, 2020). For instance, xtine burrough has examined the affective politics of the microlabor economy and the human values of the online workforce in participatory projects that subvert the MTurk's logic, such as *A Penny for Your Thoughts* (2009) in which MTurkers were paid 25 US cents to contribute a thought for a series of keywords, and *Mediations on Digital Labor* (2015) in which they were paid the same amount to do nothing for up to five minutes (burrough, 2020). Other examples include Andy Baio's *The Faces of Mechanical Turk* (2008) which tasked MTurkers to photograph themselves holding a sign on which they wrote their motivations for working on MTurk (Baio, 2008), Guido Segni's *The Middle Finger Response* (2013) in which MTurkers were paid about 50 US cents to take a webcam selfie showing their face, immediate environment, and their middle finger response (Segni, 2013), and Kieran Browne and Ben Swift's *Enacting Alternative Economies* (2020) in which MTurkers collectively authored a plain-language edition of Karl Marx's Manifesto of the Communist Party (they were paid 1.04 USD per passage) (Browne and Swift, 2020).

While some artworks realized through microlabor platforms are non-commercial, in most of them artists act as task providers, which can lead to ethically questionable situations, particularly regarding online workers' wages compared to artists' profits yielded by selling aggregately produced works (Żylińska, 2020, 117–120). The central part of this issue is that microlabor employers (task providers) define the conditions of task fulfillment and control the details about the purpose and broader context of the tasks they offer to employees. It came to the fore with Aaron Koblin's non-tactical projects *The Sheep Market* (2006), *Ten Thousand Cents* (2007–2008), and *Bicycle Built for Two Thousand* (2009, with Daniel Massey) (Koblin, 2015). Although they were conceptualized as sound generative experiments, Koblin drew critique for exploiting workers through compensatory allocation disparities (Elish, 2010; LABoral, 2011; Berdugo and Martinez, 2020, 89). For instance, in *The Sheep Market*, MTurkers were paid 2 US cents to "draw a sheep facing to the left". They generated a collection of 10,000 sheep drawings that Koblin sold online as a limited edition of adhesive stamp blocks (20 drawings/stamps per block) for 20 USD each.

Sociocultural disparities between artists and MTurkers can't be abstracted away by the wage fairness in monetary transactions, which can cause subtler ethical problems. For example, in the tactically motivated *Social Turkers* (2013), Lauren Lee McCarthy took MTurkers' real-time feedback over a mobile phone video link to improve her social interactions during twenty in-person dates she arranged on the online dating platform OkCupid (McCarthy, 2013). McCarthy was praised for positioning herself as "less than" the MTurkers by seeking their advice rather than celebrating the diversity and humanity of their responses in the face of mechanization (Berdugo and Martinez, 2020, 89). However, the notion of McCarthy's "inferiority" to the MTurkers can be interpreted as condescending because it accepts and, to some degree, conceptually exploits, even if ironically, the corporate AI's implicit categorization of MTurkers as abject (socioeconomically "low") others. Furthermore, the artist did not seek MTurkers' help as a friendly gesture (for free) but for monetary compensation—the same way the AI industry solicits their labor. In that sense, McCarthy's method for taking social and behavioral directives from the AI complex was less responsible compared to Max Hawkins' *Randomized Living* (2015–2017) which involved no monetary transaction by design. This also makes *Social Turkers* worth comparing with Stanley Brouwn's actions series *This Way Brouwn* (1961), in which the artist asked random passersby to give him verbal street directions that he audio-recorded or encouraged them to draw a navigation route to a chosen location—all voluntary exchanges in an interaction between contextually more equal actors (von Graevenitz, 1977).

---

[12] Mainstream artists soon assimilated Valla's approach. For instance, Agnieszka Kurant and John Menick's *Production Line* (2017) was a direct methodological clone of Valla's *Sequence* works, but resulted in a series of static forms (Arts at MIT 2022).

Another questionable example is Nick Thurston's *Of the Subcontract: Or Principles of Poetic Right* (2013) (Thurston, 2013) in which one MTurker was tasked to write a collection of 100 poems and a ghostwriter from Lahore, Pakistan was subcontracted to write the foreword (signed as McKenzie Wark), which were published listing Nick Thurston as the author (Wikipedia, 2022). In these cases, however, we should take into account Pablo Helguera's note (2011) that unethical artistic actions, while crossing the line of acceptability or sometimes even legality, may be defensible insofar as their (ab)uses of the expressive freedom are effective in challenging social assumptions and cultural dogmas. This is where a critical assessment that strives to be as fair and objective as possible inevitably clashes with the subtleties (and caprices) of individual ethics.

A subset of tactical practices explores AI's impact on defining, performing, and valuing (art)work by directly applying ML to art production. For instance, in *Demand Full Laziness: A Five-Year Plan* (2018–2023), Guido Segni delegated part of his artistic production to a set of GANs that make images by recording periods of his "unproductivity" such as sleeping, reading, or lazing around (Segni, 2018). Images are distributed on the Patreon platform to sponsors who fund the artist. This critique of the trend of automating artistic output integrates the contemporary reality of extreme labor extraction with the popular notions of AI-influenced cultural status of artmaking, the routinized aesthetics, and the technological opportunism of (computational) art. Similarly, Sašo Sedláček's *Oblomo* (2019–2020) contextualized the visitors' non-work with the crypto economy to satirize the NFT-driven art tokenization frenzy. It featured a blockchain web platform with custom cryptocurrency, which rewarded visitors for being still: when an ML rig detected their non-activity, it awarded them *Oblomo* coins minted from an Om block foundation. Visitors could spend these coins at the platform's market to buy services and goods offered by other visitors (Sedláček, 2020).

As Browne and Swift noted (2020, 2), computational mediation of work and creativity introduces specific concerns. It is fueled by techno-deterministic views that ignore the social contexts in which technologies are developed and applied and naturalize technologically enforced power relations. Since tactical media artists operate directly within the technological systems whose ideological frameworks they aim to critique, techno-determinism also affects the poetic, social, political, and ethical values of their work.

## Challenges

In various ways, exemplars discussed in this chapter affirm that the "Turk" in AI is stubbornly not mechanical enough, it resists "emancipation", and it is not easy to make it autonomous because humans, as inherently technological animals, are the sine qua non of technology. By facilitating the resurgence of humanity from its typically opaque habitats of computer science and industry, these artworks also assert that Silicon Valley forcefully enacts its regressively infantile dreams of ultimate techno-autonomy in the real world, with possibly nightmarish sociopolitical consequences.

At the same time, these artworks' self-critique, whether intended or not, indicates that their sociotechnical milieu simultaneously provides tactical affordances and threatens to relativize, neutralize, or instrumentalize them. Closely related to AI research, development, and application, the poetic features and implications of AI art are affected by the epistemic, conceptual, discursive, and sociopolitical problems of AI science and industry. In addition to these problems, tactical AI art extends and often amplifies tactical media art's inherent issues.

Computational art's historical tendency toward technocentrism, initially imposed by high cognitive demands of computer technology, tacitly conflates artmaking with the skillful handling of creative devices, which often incentivizes artists towards technical virtuosity devoid of self-critical distance or playful irreverence (Taylor, 2014; Żylińska, 2020, 75–85). This techno-fetishist mentality often reinforces a naïve lack of understanding that the poetic role of production techniques in the arts is fundamentally defined by conceptual thinking and meaningful contextualization. Thus, bio-detached and socially unaware notions of agency permeate AI art production and its popular representation through confused or openly mystifying rhetoric. They promote a pseudo-romantic quest for human-flavored creative "essence" within ML systems instead of demystifying them as sophisticated tools for statistical analysis, which have little to do with creativity per se and are more constructively understood in the non-technical perspective as sociopolitical apparatuses (Forbes, 2020, 6). Such notions limit the perceptiveness necessary for bold and nuanced artistic examinations of intelligence, creativity, expression, authorship, labor, ownership, authenticity, accuracy, and bias.

Various obvious but often undisclosed conceptual, methodological, thematic, aesthetic, and presentational similarities between different artists' works underscore the concerns about their creators' creative literacy and professional ethics (Grba, 2023a). Sincerely-motivated, pertinent, and well-conceived ideas are sometimes rendered as dry, unengaging, ineffective, or counter-effective works (Grba, 2022a, 20–21), represented in overly didactic setups, or supported by inflated theoretical rhetoric (Quaranta, 2020; Grba, 2021, 246–247). Tactical artworks are mostly consumed and gain significance within academic communities prone to developing echo chambers, while their real-world critical cogency, impact, and viability require unbiased assessment. The academization of tactical values (or purposes) may be scholarly fruitful, but the general audience, which is supposed to be central to tactical art, can easily recognize it as aloofness or cynicism, which leads to indifference, distrust, or resentment.

Tactical media art's sociotechnical entanglement is problematic in several aspects. By openly identifying loopholes and weaknesses in the systems they criticize, artists set their achievements up for exploitation and recuperation (Lovink and Rossiter, 2005). Sometimes, tactical critique inadvertently becomes the mere reflection of its target(s) (Raley, 2009, 30) or lapses into the mystification of technocracy whereby a class of tech-savvy artisans acts on behalf of "the (lay) people" by articulating a vision of individual freedom realizable from within the power structures of the information society (Barbrook and Cameron, 1995/2008). In the context of AI art, it is worth asking if an activist action should end up being used by the industry to enhance its profitable instruments or remedy its public image without necessarily improving its technical and ethical standards (Powles, 2018). Could tactical art disrupt the corporate AI regime with lasting and more desirable social consequences, and how effectively can it incite or aid government policies for accountability and regulation of private AI businesses with global influence?

## Perspectives

By identifying, acknowledging, and understanding these issues, artists can find new ways to intervene critically and productively in the AI-influenced sociopolitical reality. To challenge or tangibly affect AI power regimes and mobilize the audience with a lasting impact, artists need to maintain a critical outlook on their poetic devices and balance their procedural skills with motivational sincerity and ideational cogency. This ethos of balanced competencies requires cultivation through expressive diversification, experimental freedom, playfulness, conceptually strong hacking, and imaginative discovery.

Artists can benefit from epistemic humility to develop more rigorous criteria for creative thinking and acquire wider knowledge of historical, theoretical, cultural, and political contexts in which they produce and present their works (Grba, 2022a, 24). This will help them address potentially adverse expressive circumstances and clear the way for meaningful poetic directives. The inherently political nature of AI technologies (Pasquinelli, 2019) obliges artists not just to exploit them as expressive means but to recognize, deconstruct, and explore the injustices in the notional, relational, economic, and other layers of their application. To counter the recuperative sway of the corporate sector, art market, and academia, artists should strive for integrity by recognizing, objectively assessing, and correcting the systemic value issues in these domains. They need to resist prioritizing career building over artmaking, pursue external support with skepticism toward institutional rationales for art sponsorship, and be open to taking genuine risks by evolving potentially hazardous ideas.

Artists' proverbial inclination toward opportunism calls for decidedly independent professional strategies and sophisticated tactical interventions that evade the lures of commodification and complacency. These strategies require acknowledging art and technology as human dispositives within anthropological and sociocultural perspectives, and recognizing that creativity is driven by competitive ambitions and is thus inherently instrumentalizable. In that light, artists should configure and respect their methodologies as experiential processes whose outcomes inform the audience by stirring inquisitiveness and critical thinking, stimulating imagination, and encouraging progressive action. Aided by subterfuge and humor, such poetic approaches can be more impactful than superficial aestheticization, spectacularism, or overexplanation.

The contemporary AI landscape provides a range of resources for artists to reveal its sociopolitical contradictions and point out that science, technology, businesses, and education need a thorough reconfiguration and improvement of epistemological and ethical standards facing the increasing complexity of human existence. By demystifying their expressive contexts and the seemingly radical capabilities of their tools, artists can leverage AI issues as critical assets with broad political significance. The responsible treatment of these assets can build new insights about human nature and provide meaningful posthumanist perspectives (McQuillan, 2018).